\documentclass[sigconf,authorversion,nonacm]{acmart}

\usepackage{multicol}
\usepackage{multirow}
\usepackage{algorithm}
\usepackage{algorithmic}

\AtBeginDocument{%
  }

\setcopyright{acmcopyright}
\copyrightyear{2023}
\acmYear{2023}
\acmDOI{XXXXXXX.XXXXXXX}

\acmConference[KDD’24]{Proceedings of the 30th ACM SIGKDD Conference on Knowledge Discovery and Data Mining}{August 25-29, 2024}{Barcelona, Spain}
\acmPrice{15.00}
\acmISBN{978-1-4503-XXXX-X/18/06}

\begin{document}

\title{Lookahead: An Inference Acceleration Framework for Large Language Model with Lossless Generation Accuracy}

\author{Yao Zhao, Zhitian Xie, Chen Liang, Chenyi Zhuang, Jinjie Gu}
\email{nanxiao.zy, xiezhitian.xzt, liangchen.liangche, liangchen.liangche, jinjie.gujj@antgroup.com}
\affiliation{%
  \institution{Ant Group}
  \streetaddress{569 Xixi Road}
  \city{Hangzhou}
  \country{China}
  \postcode{330100}
}

% \author{Yao Zhao}
% \email{nanxiao.zy@antgroup.com}
% \affiliation{%
%   \institution{Ant Group}
%   \streetaddress{569 Xixi Road}
%   \city{Hangzhou}
%   \country{China}
%   \postcode{330100}
% }

% \author{Zhitian Xie}
% \email{xiezhitian.xzt@antgroup.com}
% \affiliation{%
%   \institution{Ant Group}
%   \streetaddress{569 Xixi Road}
%   \city{Hangzhou}
%   \country{China}
%   \postcode{330100}
% }

% \author{Chen Liang}
% \email{liangchen.liangche@antgroup.com}
% \affiliation{%
%   \institution{Ant Group}
%   \streetaddress{569 Xixi Road}
%   \city{Hangzhou}
%   \country{China}
%   \postcode{330100}
% }

% \author{Chenyi Zhuang}
% % \authornotemark[1]
% \email{c@antgroup.com}
% \affiliation{%
%   \institution{Ant Group}
%   \streetaddress{569 Xixi Road}
%   \city{Hangzhou}
%   \country{China}
%   \postcode{330100}
% }

% \author{Jinjie Gu}
% % \authornotemark[1]
% \email{jinjie.gujj@antgroup.com}
% \affiliation{%
%   \institution{Ant Group}
%   \streetaddress{569 Xixi Road}
%   \city{Hangzhou}
%   \country{China}
%   \postcode{330100}
% }

\begin{abstract}

As Large Language Models (LLMs) have made significant advancements across various tasks, such as question answering, translation, text summarization, and dialogue systems, the need for accuracy in information becomes crucial, especially for serious financial products serving billions of users like Alipay. However, for a real-world product serving millions of users, the inference speed of LLMs becomes a critical factor compared to a mere experimental model.

Hence, this paper presents a generic framework for accelerating the inference process, resulting in a substantial increase in speed and cost reduction for our LLM-based scenarios, with lossless generation accuracy. In the traditional inference process, each token is generated sequentially by the LLM, leading to a time consumption proportional to the number of generated tokens. To enhance this process, our framework, named \textit{lookahead}, introduces a \textit{multi-branch} strategy. Instead of generating a single token at a time, we propose a Trie-based retrieval and verification mechanism to be able to accept several tokens at a forward step. Our strategy offers two distinct advantages: (1) it guarantees absolute correctness of the output, avoiding any approximation algorithms, and (2) the worst-case performance of our approach is equivalent to the conventional process. We conduct extensive experiments to demonstrate the significant improvements achieved by applying our inference acceleration framework. Our framework is widely deployed in Alipay since April 2023, and obtain remarkable 2.66x to 6.26x speedup. Our code is available at https://github.com/alipay/PainlessInferenceAcceleration.

\end{abstract}


\maketitle

\section{Introduction}

Large language models (LLMs) based on transformer architecture have exhibited remarkable performance across various benchmarks, rendering them highly valuable in numerous industries. However, despite their significant achievements in language-based tasks, LLMs still face challenges in terms of inference latency when employed in generative tasks. This drawback becomes particularly apparent in scenarios where step-wise decoding is implemented. % For instance, with the GLM \cite{Du2021GLMGL} 10B model using a single Nvidia A100 GPU, the inference speed's theoretical limit through step-wise decoding is merely 93 tokens per second.

To further gain deeper insights into the factors affecting the LLMs' inference latency, we conduct a comprehensive theoretical analysis focused on a specific instance. Generating a single token with a 10B LLM (without sparse activation) necessitates reading approximately 20GB of memory and performing approximately 20G FLOPs. Although the computational load and memory access associated with the attention mechanism scale quadratically and linearly with sequence length, respectively, their impact can be considered negligible when the sequence length is less than half of the hidden dimension. For instance, in a GLM-10B model with a prompt length of 512, the memory required to read weights is 20.27GB, with additional memory for inputs and outputs totaling about 1.63GB. The computational FLOPs for matrix multiplication involving these weights are around 20.27T, whereas the combined FLOPs for attention and activation functions are approximately 0.405T. Given that an Nvidia A100 GPU provides a bandwidth of 2039 GB/s and a computational capacity of 312T FLOPs, the input/output (IO) time is estimated at 10ms (i.e., 20/2039), while the computation time is a marginal 0.06ms (i.e., 20/312,000). These calculations underscore that the IO time, rather than the computation time which is contingent on the hardware's FLOPs capacity, is the predominant factor influencing the overall inference latency of LLMs, due to its strong correlation with the model size and the available memory bandwidth.

Various techniques, such as quantization \cite{Lin2023AWQAW,Frantar2022GPTQAP}, sparsity \cite{Lu2023STEPLN,zhou2021learning}, pruning \cite{zhang2022platon,lagunas2021block},  distilling \cite{jeon2023pet,hsieh2023distilling}, and tensor decomposition \cite{ma2019tensorized,wang2023tensor}, have been proposed to reduce the LLMs' size and the IO consumption time for predicting each token in LLMs. However, these techniques have been found to result in a degradation of accuracy. To address the challenge of predicting more tokens within a fixed IO consumption time, non-autoregressive neural machine
translation (NAT) \cite{Gu2017NonAutoregressiveNM} and iterative parallel decoding \cite{santilli2023accelerating} have been introduced and successfully applied in translation tasks. Unfortunately, this approach has shown limited effectiveness in question-answering scenarios. 

Recently, speculative decoding with a draft model has become a popular strategy. However, this strategy necessitates extra training efforts \cite{blockdecoding,cai2024medusa,robustexit} or a smaller auxiliary model capable of producing  drafts \cite{xia2023speculative, chen2023accelerating, miao2024specinfer}, and they may worse the memory burden with additional parameters or models. In light of this, training and assist-model free strategies are proposed, such as LLMA \cite{Yang2023InferenceWR} and LookaheadDecoding \cite{fu2023lookahead}. The LLMA algorithm relies on a text-matching technique, which is effective within document retrieval domains, tends to underperform in other applications. LookaheadDecoding, on the other hand, incorporates the Jacobi iteration for draft generation, which, despite its innovative design, may face computational bottlenecks that impair its overall effectiveness.

Table \ref{tbl.comparison} summarizes the acceleration techniques discussed previously. To address their limitations, we introduce \textit{Lookahead}, a groundbreaking framework that incorporates a trie-tree-based retrieving strategy and a multi-branch-based parallel Verification and Accept (VA) strategy. \textit{Lookahead} leverages a trie tree to record the n-gram tokens of input prompts and generated responses. The draft is retrieved based on the provided context tokens, allowing for extremely fast draft generation. The draft tokens arranged in a logical tree structure are efficiently processed in the Verification and Accept (VA) process. Furthermore, we have implemented an adaptive strategy to optimize the retrieval process, effectively striking a balance between memory and computation requirements. It shows that \textit{Lookahead} proves its superior performance in accelerating LLMs' inference, compared with the existing state-of-the-art (SOTA) acceleration method.

\begin{table}[ht]
\caption{Comparison of different acceleration methods.}
\small
\setlength\tabcolsep{2pt}
\label{tbl.comparison}
\begin{tabular}{ccccc}
\toprule  
% \cline{2-5}
\multirow{3}{*}{Methods} &  Accuracy- & Training-  & Multi- & Low-cost- \\ 
 &  lossless  & or-Assis- &  branch- & draft- \\ 
 &  & Model-free  & draft & generation \\ 

\midrule  
Quantization \cite{Lin2023AWQAW,Frantar2022GPTQAP} & $\times$  & - & - & - \\
Sparsity \cite{Lu2023STEPLN,zhou2021learning} & $\times$   & - & - & -  \\
Pruning \cite{zhang2022platon,lagunas2021block} & $\times$   & - & - & -  \\
Distilling \cite{jeon2023pet,hsieh2023distilling} & $\times$ & $\times$ & -  & -  \\
Tensor-decomp \cite{ma2019tensorized,wang2023tensor} & $\times$ & $\times$ & -  & - \\
Early-exit \cite{li2021accelerating,xin2020deebert} & $\times$ & $\times$ & -  & - \\

\midrule  

Block decoding \cite{blockdecoding} & \checkmark & $\times$ & $\times$  & \checkmark \\
Spec decoding \cite{xia2023speculative} & \checkmark  & $\times$ & $\times$  & $\times$ \\
SpecInfer \cite{miao2024specinfer}  & \checkmark & $\times$ & \checkmark   & $\times$ \\
FREE \cite{robustexit} & \checkmark & $\times$ &  $\times$  & \checkmark \\
LLMA \cite{Yang2023InferenceWR} & \checkmark & \checkmark & $\times$   & \checkmark \\
LookaheadDecoding \cite{fu2023lookahead} & \checkmark & \checkmark & \checkmark   & $\times$  \\
Lookahead (ours) & \checkmark & \checkmark & \checkmark   & \checkmark \\
\bottomrule 
\end{tabular}
\end{table}

Being benefited from the superior performance and accessibility, our \textit{Lookahead} framework has been widely employed in dozens of the real world scenarios of Alipay, including financial RAG, health suggestion, medical report summary, etc.

Moreover, to gain a wide range of applications, we have implemented our framework based on the transformers library of Hugging face\footnote{https://huggingface.co/}, by extending a generation mode named \textit{lookahead generation}, which supports the \textit{greedy search} and \textit{sample} generation strategy. We have also currently applied \textit{Lookahead} to the most recent LLMs, such as GLM\cite{Du2021GLMGL}, Llama\cite{touvron2023llama}, OPT\cite{Zhang2022OPTOP}, GPT2\cite{Radford2019LanguageMA}, BLOOM\cite{Scao2022BLOOMA1}, ChatGLM\cite{Zeng2022GLM130BAO},  Baichuan\cite{Yang2023Baichuan2O} and Qwen\cite{Bai2023QwenTR}, InternLM\cite{2023internlm}, Mistral\cite{jiang2023mistral}, Mixtral MoE\cite{jiang2024mixtral}, etc. The aforementioned models can be easily adapted to integrate \textit{Lookahead}, our well-designed framework, with only minor code modifications of approximately 20 lines, which can be found in our repository.

Our contributions can be summarized as:
\begin{itemize}
    \item We empirically quantify that the main bottleneck of LLM inference is the IO bandwidth, rather than the computation bound.
    \item We innovatively develop \textit{Lookahead}, a framework that applies a hierarchical multi-branch draft strategy implemented with a trie tree to output more tokens per step than the traditional methods, in accelerating LLMs' inference.
    \item We extensively conduct experiments on both the industry and open source datasets and prove that \textit{Lookahead} brings a significant improvement over the existing SOTA method in accelerating LLMs' inference.
    \item We elaborately adapt \textit{Lookahead} to the most recent LLMs without any assistance of smaller models and have released our work with open source.
\end{itemize}

\section{Related Work}

Recently, several strategies have been proposed and developed to enhance the inference speed of LLMs while maintaining the output quality within an acceptable range. One such strategy is the non-auto-regressive approach, specifically non-auto-regressive translation (NAT) \cite{Gu2017NonAutoregressiveNM}, primarily employed in translation tasks \cite{kasai2021deep, Huang2021NonAutoregressiveTW, saharia-etal-2020-non}. However, it is essential to note that there are significant distinctions between translation tasks and general language model (LLM) scenarios, which may lead to subpar performance when applying the NAT strategy to LLM decoding. 

To address this limitation, Huang et al. \cite{Huang2021NonAutoregressiveTW} introduce a layer-wise iterative method wherein each layer utilizes the decoding results and embeddings from the preceding layers. This approach involves training each layer through maximum likelihood estimation to predict the outcomes of subsequent decoding layers. On the other hand, Santilli et al. \cite{santilli2023accelerating} formalized the standard greedy auto-regressive decoding strategy by employing a parallel Jacobi and Gauss-Seidel fixed-point iteration. It initializes the next tokens using special tokens and performs iterative decoding until convergence. However, all these methods may suffer from the risk of accuracy degeneration, since the manipulated model deviates from its original version.

Recently, there has been a proposal for accuracy-lossless acceleration to enhance the auto-regressive decoding strategy as illustrated in Table \ref{tbl.comparison}. In this approach, a block-wise parallel decoding strategy was introduced by \cite{blockdecoding}. In this strategy, each subsequent token is independently and parallel predicted as a proposal using an additional transformer model, which consists of a multi-output feed-forward layer and can be fine-tuned or distilled for optimal performance. Then the proposals are directly compared against the output tokens generated by the original decoder. The longest verified tokens are then selected for acceptance as the current step's output. However, frequent failures during the verification process may be occur due to its reliance on a singular predictive branch. To overcome this drawback, Medusa \cite{cai2024medusa} employs multiple heads to simultaneously predict drafts, enhancing the robustness of the process. On the other hand, FREE \cite{robustexit}, uses the shallow layers of a model to generate drafts, instead of the final layer. to address this dependency. It proposals a synchronized parallel decoding strategy to ensure accuracy without loss. Despite this strategy's remarkable acceleration, more efforts are needed to train the extra layer.

To address the aforementioned issue, speculative decoding has been proposed  \cite{xia2023speculative, chen2023accelerating, miao2024specinfer, zhang2024accelerating, hong2024simple}. These works utilize a smaller model as a draft predictor. For instance, the Bloom 7.1B model can be employed as a draft model for the larger 176B model. However, the proposed works above may face significant practical challenges. Primarily, the availability of a smaller-scale model, typically one-tenth the size of the larger model, is not always feasible for a range of model series, such as Llama-7B \cite{touvron2023llama} and ChatGLM-6B \cite{Zeng2022GLM130BAO}. Furthermore, larger models are often fine-tuned to specific applications, necessitating a parallel fine-tuning of the smaller helper model to produce comparable drafts. This requirement complicates the deployment and diminishes the system's accessibility.
% Recently, Miao et al \cite{miao2024specinfer} introduces SpecInfer, a system that accelerates generative LLM by leveraging small speculative models to predict the LLM's outputs.

% However, this system may meet serious limitations in practical scenarios. Firstly, a smaller model may be not available, usually we need the small model to be tenth scale of the large model to obtain notable acceleration, which is difficult for a series models (e.g., Llama-7B \cite{touvron2023llama}, ChatGLM-6B \cite{Zeng2022GLM130BAO}). Moreover, a smaller model may have diverse output to the large model, as the large models usually be fine-tuned to adapt to the practical scenarios, one has to fine-tuned the assist model to generate similar drafts for the large model, which also deteriorates accessibility.

% However, the proposed works above may face significant practical challenges. Primarily, the availability of a smaller-scale model, typically one-tenth the size of the larger model, is not always feasible for a range of model series, such as Llama-7B \cite{touvron2023llama} and ChatGLM-6B \cite{Zeng2022GLM130BAO}. Furthermore, larger models are often fine-tuned to specific applications, necessitating a parallel fine-tuning of the smaller helper model to produce comparable drafts. This requirement complicates the deployment and diminishes the system's accessibility.

To overcome this challenge, model-free prediction strategies have been introduced to achieve accurate predictions without relying on a specific model. One such strategy is presented in Ge et al. \cite{ge2022lossless}, which utilizes an input-guided method that copies content from the input sentence through prefix matching. Another strategy LLMA, proposed by Yang et al. \cite{Yang2023InferenceWR}, employs a prefix matching approach to retrieve content from either the input sentence or a document database. However, it is worth noting that the model-free prediction strategies mentioned above utilize tokens in a single draft manner, failing to fully utilize the GPUs. Recently, another training-free and assist-model-free method named LookaheadDecoding \cite{fu2023lookahead} explores the multi-branch strategy with employing Jacobi iteration and speculative decoding simultaneously. Nonetheless, this method incurs a substantial overhead associated with the generation of drafts, which consequently attenuates the potential for acceleration.

\section{Preliminary}

\subsection{Inference Speed}
The inference speed \textit{V} can be expressed as below.

\begin{equation}
\begin{aligned}
    V = \frac{L}{T} \propto \frac{L}{N \times t(l)} 
\end{aligned}
\label{eq:position_func}
\end{equation}

\begin{equation}
\begin{aligned}
    N \propto \frac{L}{l} 
\end{aligned}
\label{eq:position_func}
\end{equation}

Here, \textit{L} denotes the overall generation tokens' length and \textit{T} is the overall inference time, which is positively correlative to the overall consuming time for decoding: \textit{N} indicates the overall decoding steps, $t(l)$ is the decoding time per step. It should be noted that $t(l)$ is nearly constant while \textit{l}, namely the generated tokens' length per decoding step, is within a certain range, whose details and explanations will be introduced and discussed in the following sub-section. Therefore, given the fixed \textit{L}, the longer \textit{l} is, the fewer \textit{N} is needed, which in turn promises a higher inference speed \textit{V}.

\subsection{Step-Wise Decoding}
Auto-regressive language models have been firstly introduced through following a step-wise decoding strategy: at each decoding step, the models concatenate the prompt sequence and the previous generated tokens and output the next single token using greedy-decoding, which selects the next token with the highest predicted probability over the vocabulary. 

Though this strategy has been widely applied, the particular process promises only one single output token per decoding step ($l=1$), which limits the overall inference speed.

\subsection{Singe-Branch Strategy}
Several most recent methodologies \cite{blockdecoding} \cite{Yang2023InferenceWR} have been proposed to generate a sequence of tokens at each decoding step, with the purpose of promising a higher \textit{l} to accelerate the LLMs' inference speed \textit{V}. In these works, according to the prompt and the output tokens at the previous decoding step, a branch of tokens, named single-branch draft, have been obtained through the small model or the document reference, and efficiently validated by running the LLM in a single forward process. 

Despite the single-branch strategy's success in accelerating LLMs' inference, the generated tokens' length per decoding step, namely \textit{l}, cannot be guaranteed as long as we wish. The previous works \cite{blockdecoding} \cite{Yang2023InferenceWR} empirically conclude that there is an upper limit for \textit{l} and the inference speed \textit{V}. 

This can be explained through the successive validating mechanism: the validation process breaks once one token in the single-branch draft fails the validation, only the validated tokens in front of this failed token are accepted as the output. For the sake of brevity, we give a definition below:

\textbf{\textit{Definition 1}} Following the predicted next token, only the successive validated tokens from the beginning of the branch draft are kept and accepted as the output tokens, whose length is called the \textbf{\textit{effective decoding length, EDL}}. 

Simply extending the single-branch draft's length over \textbf{\textit{EDL}} not only fails to promise a longer output tokens' length per decoding step \textit{l} and a higher inference speed \textit{V}, but also wastes more computation. Apparently, how to achieve a longer \textbf{\textit{EDL}}, namely the \textbf{\textit{effective decoding length}}, is the key to further accelerate LLMs' inference.

\subsection{Discovering GPU's FLOPs Redundancy}
To explore how far \textbf{\textit{EDL}} can be extended at each decoding step, we conduct a case study through discussing the decoding length's impact on the overall consuming time of LLMs' single forward process, whose result can be found in Figure 1. It should be noted that we apply AntGLM-10B (which is trained from scratch with the GLM structure \cite{Du2021GLMGL,Zeng2022GLM130BAO}) model with single Nvidia A100 GPU as an example. 

\begin{figure}
    \centering
    \includegraphics[width=3.0in]{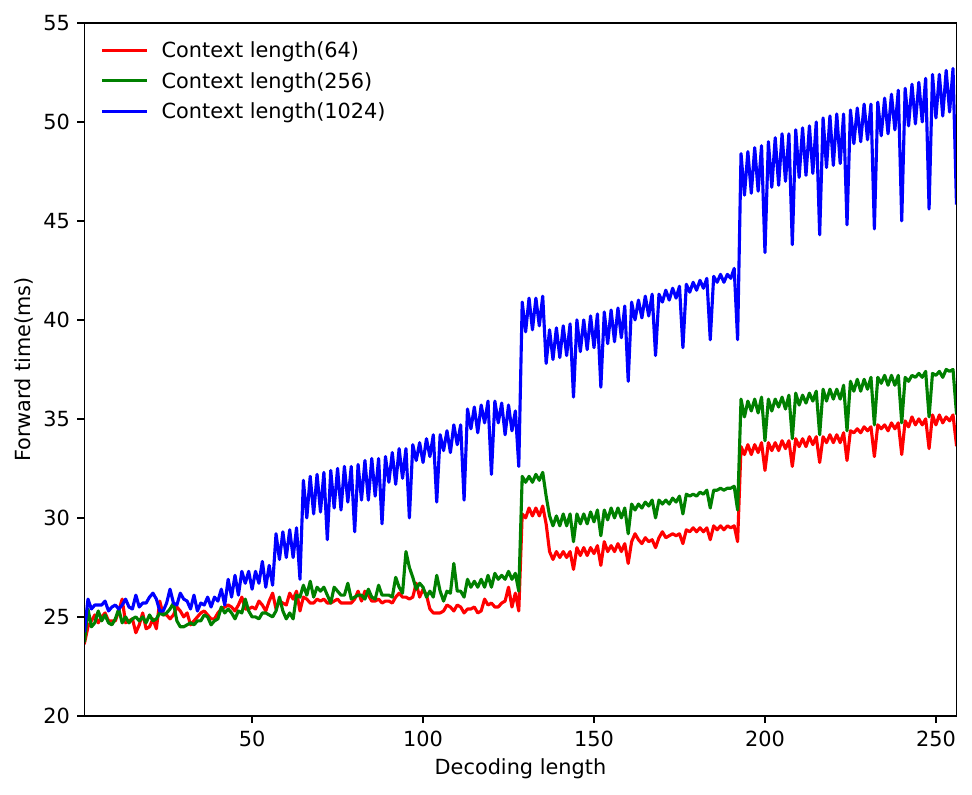}
    \caption{Decoding length's impact on the overall consuming time of LLMs' single forward process. Even the forward FLOPs is linear to the decoding length,}
    \label{fig:batch}
\end{figure}

In Figure \ref{fig:batch}, it is evident that as the decoding length increases in the LLM's single forward process, the time consumption remains relatively constant in the initial stage, given a fixed context length. This observation can be attributed to several factors. Firstly, certain time requirements, such as kernel launching, operator loading, reading, and verification, are fixed and dependent on the GPU bandwidth. These factors are unrelated to the computational complexity or decoding length. Secondly, when the decoding length is relatively short, the resulting small matrix blocks may not take full advantage of CUDA and Tensor cores, indicating that GPU shows its FLOPs redundancy in this scenario. Consequently, the differences in overall time consumption between decoding lengths of 16 and 128 in the LLM's single forward process is negligible, provided the context length is 256. This finding supports the concept that the decoding time per decoding step, represented as \textit{t(l)}, remains nearly constant within a specific range of generated token lengths \textit{l}. However, when the decoding length exceeds 128 while the context length is 256 in the LLM's single forward process, the larger matrix block size introduces a more complex calculation logic for the tensor cores. Consequently, the overall inference time shifts to the second stage, leading to a gradual increase in overall inference time.

\textbf{\textit{Definition 2}} We define the maximum decoding length within which the inference consuming time of the single forward process is nearly constant as the \textbf{\textit{critical decoding length, CDL}}.

It can be clearly seen that there is still a "sufficient gap" between \textbf{\textit{EDL}}s empirically concluded previously \cite{blockdecoding} \cite{Yang2023InferenceWR} and the \textbf{\textit{CDL}} illustrated in Figure 1. In this scenario, we come up with two questions:

1. How to extend \textbf{\textit{EDL}} further to accelerate the LLMs' inference as much as possible?

2. How to fully leverage the capabilities of GPUs, particularly considering our preliminary finding that GPUs exhibit FLOPs redundancy when the decoding length is within the \textbf{\textit{CDL}}?

Motivated by these questions, we propose a framework called \textit{Lookahead} that aims to achieve a longer \textbf{\textit{EDL}} by optimizing GPU utilization. The subsequent section provides detailed insights into this framework.

\section{Methods}
\subsection{Overview of Lookahead}
We construct a framework, \textit{Lookahead}, to accelerate LLMs' inference through developing and applying the \textit{multi-branch strategy}.

In contrast to the single-branch strategy such as \cite{Yang2023InferenceWR}, which only considers one available draft without considering the existence of other drafts, the multi-branch strategy retrieves multiple drafts simultaneously. These drafts are then efficiently decoded and validated in parallel through the Verification and Accept (VA) process, progressively. The VA process then identifies the correct sub-sequence for each draft and retains the longest sub-sequence as the output tokens. Figure 2 presents the progress overview including various draft retrieving and verifying strategies.

The execution of the multi-branch strategy is based on a fundamental fact that when considering a sequence of tokens, there may exist multiple sequences of successive following tokens, each representing a potential branch draft. These sequences are referred to as \textit{multi-branch draft} in our work, whose details will be introduced in the next sub-section.

\begin{figure*}
    \centering
    \includegraphics[width=0.9\linewidth]{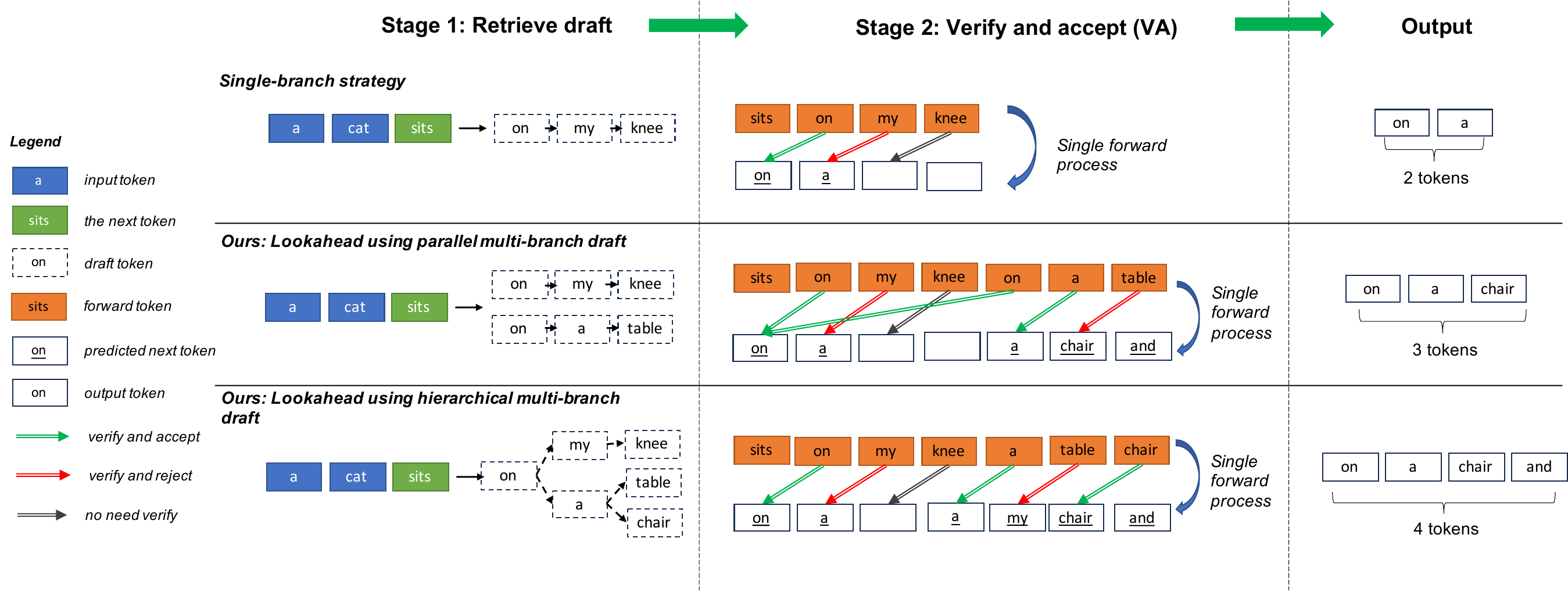}
    \caption{Overview of the drafts retrieving and the Verification and Accept (VA) process using various strategies.}
    \label{fig:draft}
\end{figure*}

\subsection{Multi-Branch Draft}
While developing and applying the multi-branch strategy, a concern naturally arises regarding how to efficiently deal with as many drafts as possible within the constraints of limited computational resources, specifically the \textbf{\textit{CDL}}.

\subsubsection{Parallel Multi-Branch Draft}
The drafts organized in a simple and straight forward parallel manner are referred to as \textit{parallel multi-branch draft}, whose details can be found in Figure 2 and 3.

\subsubsection{Hierarchical Multi-Branch Draft}
Fortunately, through careful observation of the context, we have noticed that certain branches in the drafts have common prefix tokens. For example, in Figure 2, the branches \textbf{[on, my, knee]} and \textbf{[on, a, table]} both share the prefix token \textbf{[on]}. By leveraging this observation, we can compress and organize the multi-branch draft using a hierarchical structure, allowing for the inclusion of additional branches while maintaining the same token capacity. To achieve this, we recursively merge the prefix token(s) shared by multiple branches, and then append the remaining tokens of each branch. We refer to these compressed and organized drafts as \textit{hierarchical multi-branch draft}. 

As illustrated in Figure 2 describing the process of an inference step, through the single-branch strategy, the single draft \textbf{[on, my, knee]} is appended to the next token \textbf{[sits]} directly for the single forward process in VA, which in turn gives the output \textbf{[on, a]}. Being benefited from the parallel multi-branch draft, two branches \textbf{[on, my, knee]} and \textbf{[on, a, table]} are organized in parallel and appended to the next token \textbf{[sits]} for verifying, which outputs \textbf{[on, a, chair]}. By utilizing the hierarchical multi-branch draft, the shared prefix token \textbf{[on]} are merged, after which the token list \textbf{[my, knee, a, table]} are appended. By doing so, given the same decoding length, we are able to save an additional space to append one more token \textbf{[chair]} from another branch \textbf{[on, a, chair]}. Combined with the next token, we collect the token list \textbf{[sits, on, my, knee, a, table, chair]} for decoding through the single forward process and output \textbf{[on, a, chair, and]}. Therefore, compared to the single-branch strategy, our multi-branch strategy, particularly the hierarchical multi-branch draft, offers the advantage of retrieving multiple drafts, resulting in improved \textbf{\textit{EDL}} from a statistical perspective and thus significantly enhancing LLMs' inference speed. As illustrated in Figure 3, the position IDs and causal masks in a transformer-based model \cite{Vaswani2017AttentionIA} are also merged to align with the merged token list. By doing so, we are able to accommodate more branches using \textit{hierarchical multi-branch draft}, compared to \textit{parallel multi-branch draft}.

\begin{figure*}
    \centering
    \includegraphics[width=6.5in]{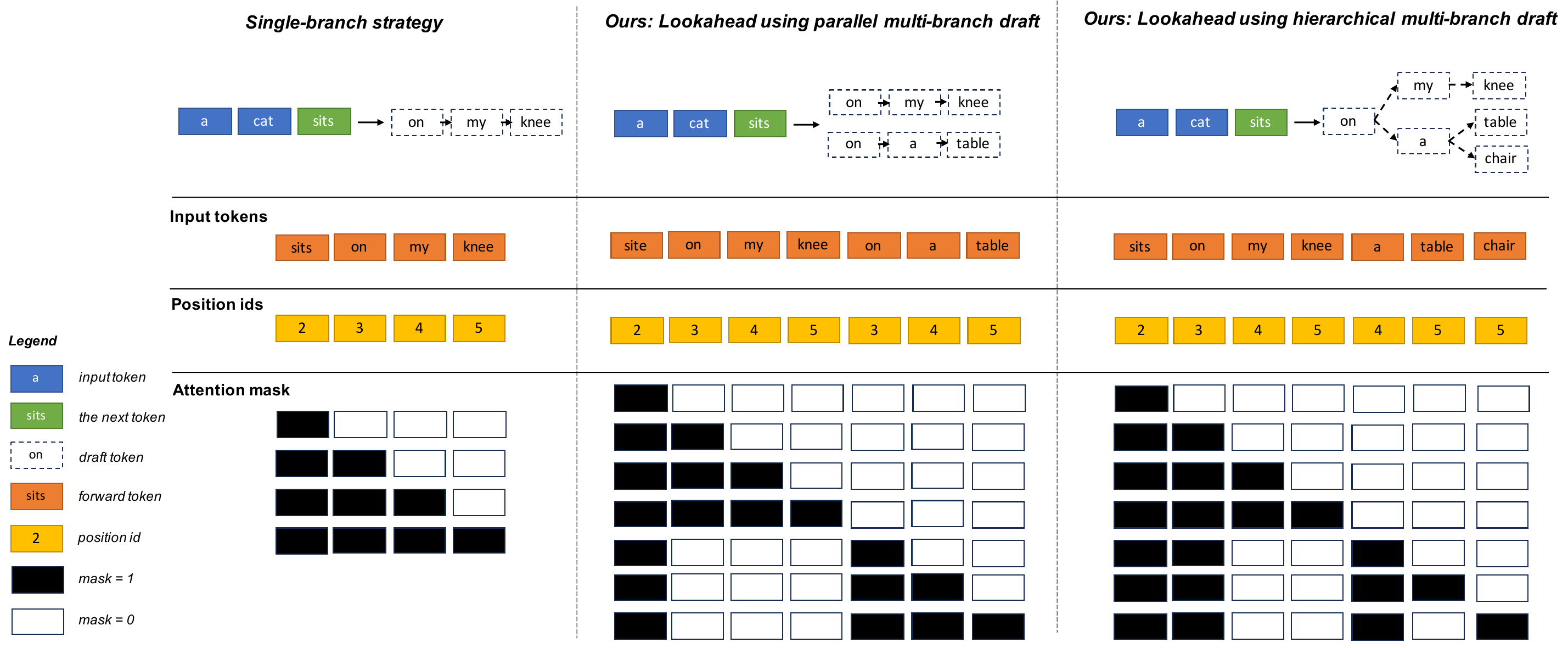}
    \caption{The input ids, position ids and causal masks for forwarding using various strategies.}
    \label{fig:mask}
\end{figure*}

\subsection{Trie-tree-based Draft Retrieval}
To enable \textit{hierarchical multi-branch draft}, we utilize a trie tree \cite{Briandais1959FileSU, 10184017}, a data structure that is widely used for efficient retrieval and storage purposes and handles prefix matching by organizing nodes as individual characters or words. In our work, each node of the trie tree represents a token ID, and a path from the root to a leaf node represents a branch. The trie tree is initialized when a model is loaded, and it is alive until the model instance is shut down. 

Before and after each step consisting of the draft retrieving process and the VA process as illustrated in Figure \ref{fig:draft}, a global trie tree will be updated through multiple procedures, which will be introduced in the sub-section \ref{trie-tree-update}. During the draft retrieving process, the trie tree will be retrieved to provide the drafts, whose details will be introduced in the sub-section \ref{trie-tree-retrieve}.

\subsubsection{Trie Tree Updating}\label{trie-tree-update}
We apply the branch inserting, branch eliminating and node pruning to update the trie tree.

\textbf{Prompt branch Inserting.} Considering that in several typical scenarios (e.g., RAG, Summary), the output are often derived from the prompt, we thus manipulate the prompt into the branches and insert them into the trie tree. 

\textbf{Generated branch Inserting.} Different from LLMA which only uses branches from prompts, we also  reuse the branch from generated tokens. We find that, towards various prompts, there may be similar outputs, e.g., "keep exercising" in answering "how to keep healthy" and "how to be stronger". Moreover, we also find that some tokens may be repeated in a response, from example, "numbers.push\_back" may occur several times in a code-generation response. To benefit from the repeat, we put the generated branches into the trie tree on-the-fly instead of the final step.

\textbf{Branch Eliminating.} When the current generation in answering the prompt is finished after multiple steps, the branches derived from this prompt are eliminated, considering that these branches may not be relevant to the generation in answering the other prompts.

\textbf{Node Pruning.} To maintain the trie tree within a moderate size, in case that the oversized trie tree results in high memory consumption and slow retrieval performance, we develop a pruning strategy. To achieve this, we decay the branch frequency and remove the nodes with frequency less than 1, when the trie tree exceeds a predetermined capacity. By doing so, we are able to optimize memory consumption and improve retrieval performance in our framework.

The analysis of the above updating procedures will be discussed in the experiment section.

\subsubsection{Trie Tree Retrieving}\label{trie-tree-retrieve}

\textbf{Multi-stage Retrieval.} We can extract a sub tree from a trie tree by providing a prefix, which is essentially a list of tokens. The sub tree is also a trie tree and can be directly used for the hierarchical multi-branch draft. The length of the prefix influences the number and relevance of the retrieved branches. Short prefixes yield a greater number of branches, while longer prefixes result in more related branches. To strike a balance between the count and correlation of the branches, we adopt a multi-stage retrieval strategy inspired by \cite{Yang2023InferenceWR}. Specifically, we begin by attempting to match a longer prefix. If the number of tokens associated with the matched branches is significantly smaller than the \textbf{\textit{CDL}}, we reduce the length of the prefix and retry the matching process until we obtain a substantial number of tokens linked to matched branches. If the count of the matched branches falls below a predefined threshold, we utilize all of them for the VA process. However, if the count exceeds a given size, we choose the tokens with the highest frequency. 

\textbf{Branch Weighting.} For token sorting by frequency, we employ a weighted scheme that considers both prompt and response frequencies.  Intuitively, branches from an input prompt may be more related to current generation than that of other responses. To prioritize the importance of branches derived from the input prompt, we amplify their frequency by a significant factor.

\section{Experiments}

\subsection{Experimental Setup}
We conduct a solid experiment to evaluate the efficacy, robustness and practicality of our \textit{Lookahead} framework in accelerating LLMs' inference. Our inference environment includes a server with 32-core CPU, 64 GB host memory, and various GPUs.
Considering the actual industry scenarios that \textit{Lookahead} is applied, AntRAG is chosen as the evaluation dataset. AntRAG dataset is an internal benchmark dataset that is meticulously collected from a real-life product system in Alipay. The dataset collection process involves submitting user queries to a search engine, which then retrieves the most relevant document alongside the original query. These query-document pairs are combined to form prompts. To ensure the dataset's quality and coherence, we employ a rigorous process of removing any duplicate or unrelated prompts, resulting in a refined and comprehensive dataset. Furthermore, to validate the robustness and practicality of \textit{Lookahead} in other scenarios, we additionally conduct experiments utilizing the Dolly dataset (an open domain QA dataset) \footnote{https://huggingface.co/datasets/databricks/databricks-dolly-15k}, GSM8k dataset (a dataset of 8.5K high quality linguistically diverse grade school math word problems) \footnote{https://huggingface.co/datasets/gsm8k}, and HumanEval-x dataset (a dataset of Python, C++, Java, JavaScript, and Go code tasks) \footnote{https://huggingface.co/datasets/THUDM/HumanEval-x}. Detailed information regarding the datasets, base models and devices utilized can be found in the Appendix. For inference speed evaluation, each dataset's test set is utilized, with the corresponding development set employed for warm-up purposes. It is important to clarify that this warm-up procedure is implemented solely to ensure the convergence and accuracy of performance metrics, as detailed in the Appendix, and is not included in real-world deployment scenarios.
In evaluating the candidate methods, which all purport lossless accuracy in generation, inference speed, quantified as output tokens generated per second, serves as the sole metric. Optimal hyper-parameters are determined through a grid search, with the grid size deliberately set as multiples of four to enhance time efficiency.

\subsection{Results}

\subsubsection{Lookahead's superior performance}
Table \ref{tbl.speed_summary} exhibits the inference speeds achieved by different acceleration methods. The mean value of the inference speed is then selected as the performance indicator for each acceleration method.

% \begin{table}[htb]
% \caption{Inference Speed of different methods and datasets. The results are calculated by conducting five repeated runs.}
% \small
% % \setlength\tabcolsep{3pt}
% \label{tbl.speed_summary}
% \begin{tabular}{ccccc}
% 	\toprule  
% \multirow{2}{*}{Methods} & \multicolumn{2}{c}{AntRAG} &  \multicolumn{2}{c}{Dolly} \\ 
% \cline{2-5}
%  & token/s & speedup  & token/s & speedup  \\ 
% \midrule  
% Baseline & 52.4±0.2 & 1.00 & 34.0±0.3 & 1.00 \\
% LLMA  & 165.4±3.3 & 3.16 & 50.8±0.2 & 1.49 \\
% Lookahead (Parallel) & 263.4±3.3 & 5.03 & 68.9±0.4 & 2.03  \\
% Lookahead (Hierarchical) & \textbf{280.9±4.4} & \textbf{5.36} & \textbf{71.7±0.9} & \textbf{2.11} \\
% \bottomrule 
% \end{tabular}
% \end{table}

\begin{table*}[htb]
\caption{Inference Speed(token/s) of different methods on various models, datasets and devices. As LookaheadDecoding only offers implementation for Llama, we do not conduct experience on AntGLM-10B Model.}
% \small
% \setlength\tabcolsep{3pt}
\label{tbl.speed_summary}
\begin{tabular}{cccccccc}
\toprule  
model & dataset & device & transformers & vLLM & LLMA & LaDe & lookahead(ours)  \\
\midrule  
AntGLM-10B & AntRAG & A100-80G & 52.4 & 52.06(x0.99) &  165.4(x3.16) & - & 280.9(x5.36) \\
AntGLM-10B & AntRAG & A10 & 20.3 & 20.29(x1.00) &  59.5(x2.93) & - & 105.1(x5.18) \\
AntGLM-10B & AntRAG & V100-32G & 27.3 & 27.28(x1.00) &  64.3(x2.36) & - & 118.9(x4.36) \\
Llama-7B & Dolly & A100-80G & 50.4 & 91.04(x1.81) &  60.7(x1.20) & 66.7(x1.32) & 106.8(x2.12) \\
Llama-7B & Dolly & A10 & 31.4 & 32.58(x1.04) &  35.8(x1.14) & 38.2(x1.22) & 55.7(x1.77) \\
Llama-7B & GSM8k & A100-80G & 41.4 & 92.09(x2.22) &  69.0(x1.67) & 94.3(x2.28) & 111.3(x2.69) \\
Llama-7B & GSM8k & A10 & 31.4 & 32.73(x1.04) &  41.1(x1.31) & 48.9(x1.56) & 68.1(x2.17) \\
Llama-7B & HumanEval-x & A100-80G & 51.1 & 90.47(x1.77) &  66.5(x1.30) & 88.3(x1.73) & 161.5(x3.16) \\
Llama-7B & HumanEval-x & A10 & 30.9 & 32.46(x1.05) &  42.1(x1.36) & 46.0(x1.49) & 89.6(x2.90) \\
Llama-13B & Dolly & A100-80G & 39.9 & 51.67(x1.29) &  59.0(x1.48) & 45.2(x1.13) & 84.6(x2.12) \\
Llama-13B & Dolly & V100-32G & 20.5 & 22.07(x1.08) &  23.9(x1.17) & 23.2(x1.13) & 35.2(x1.72) \\
Llama-13B & GSM8k & A100-80G & 42.9 & 52.06(x1.21) &  51.8(x1.21) & 64.8(x1.51) & 103.4(x2.41) \\
Llama-13B & GSM8k & V100-32G & 22.0 & 22.43(x1.02) &  25.8(x1.17) & 27.8(x1.26) & 45.6(x2.07) \\
Llama-13B & HumanEval-x & A100-80G & 35.0 & 51.49(x1.47) &  57.7(x1.65) & 66.1(x1.89) & 137.3(x3.92) \\
Llama-13B & HumanEval-x & V100-32G & 21.5 & 22.33(x1.04) &  28.8(x1.34) & 27.5(x1.28) & 57.0(x2.65) \\
\bottomrule 
\end{tabular}
\end{table*}

As depicted in Table \ref{tbl.speed_summary}, our \textit{Lookahead} obtain significant improvement over other methods, on various models, datasets and devices. The average inference speed of AntGLM-10B towards AntRAG is recorded at 52.4 tokens/s on A100 GPU. However, with the incorporation of LLMA, this speed is elevated to 165.4 tokens/s, resulting in a notable 3.16 times speed improvement. Our \textit{Lookahead} with hierarchical multi-branch draft via trie tree propels the average inference speed even further, culminating in a 5.36 times speed-up, surpassing LLMA by 70\% in terms of acceleration. \textit{Lookahead} also obtains similar acceleration with AntRAG dataset on less-powerful devices, such as A10 and V100, which are widely used for deployment. Despite of RAG scenarios, \textit{Lookahead} also consistently demonstrates its remarkable superiority and applicability across diverse datasets, especially with speedup of 3.92 on the HumanEval-x dataset, further emphasizing its practicality in real-world scenarios.

\subsubsection{Hyper-parameters in multi-branch draft}
\begin{figure}[htb]
	\centering
	\includegraphics[width=3.0in]{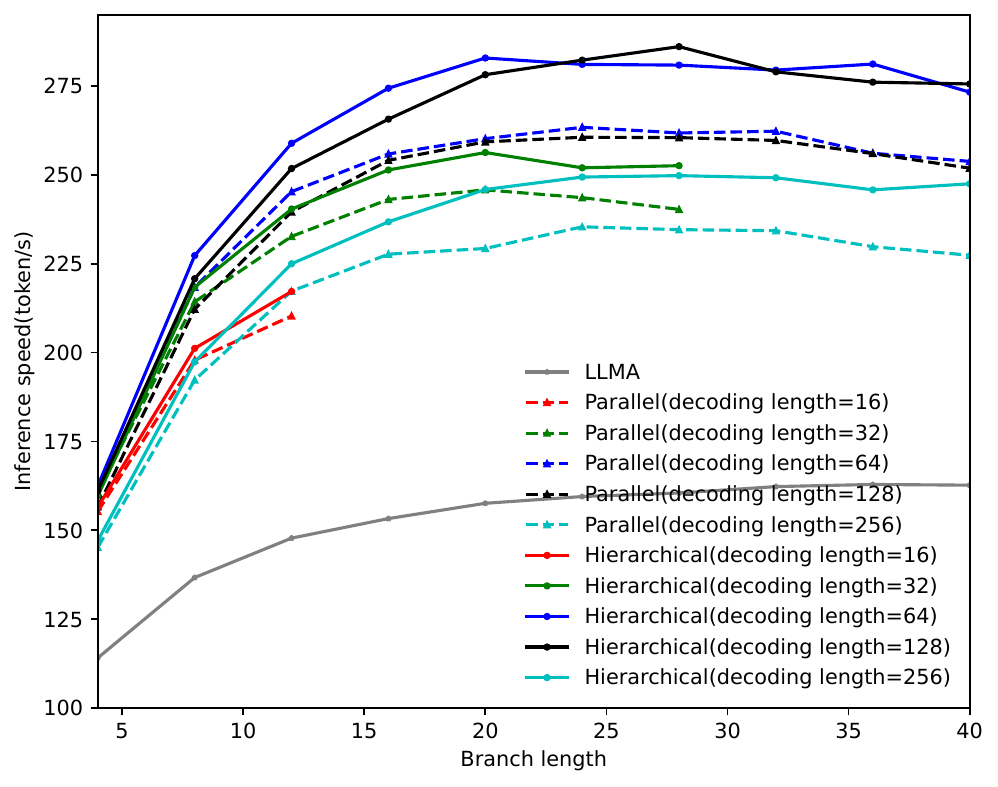}
	\caption{The decoding and branch length's impact on the LLM's inference speed using various accelerations.}
	\label{fig.speed}
\end{figure}

In continuation of AntGLM-10B, we delve deeper into the analysis of the decoding and branch lengths, two key hyper-parameters within our \textit{Lookahead} framework. In particular, we empirically examine their impact on the inference speed, as illustrated in Figure \ref{fig.speed}.

\begin{figure}[htb]
	\centering
	\includegraphics[width=3.0in]{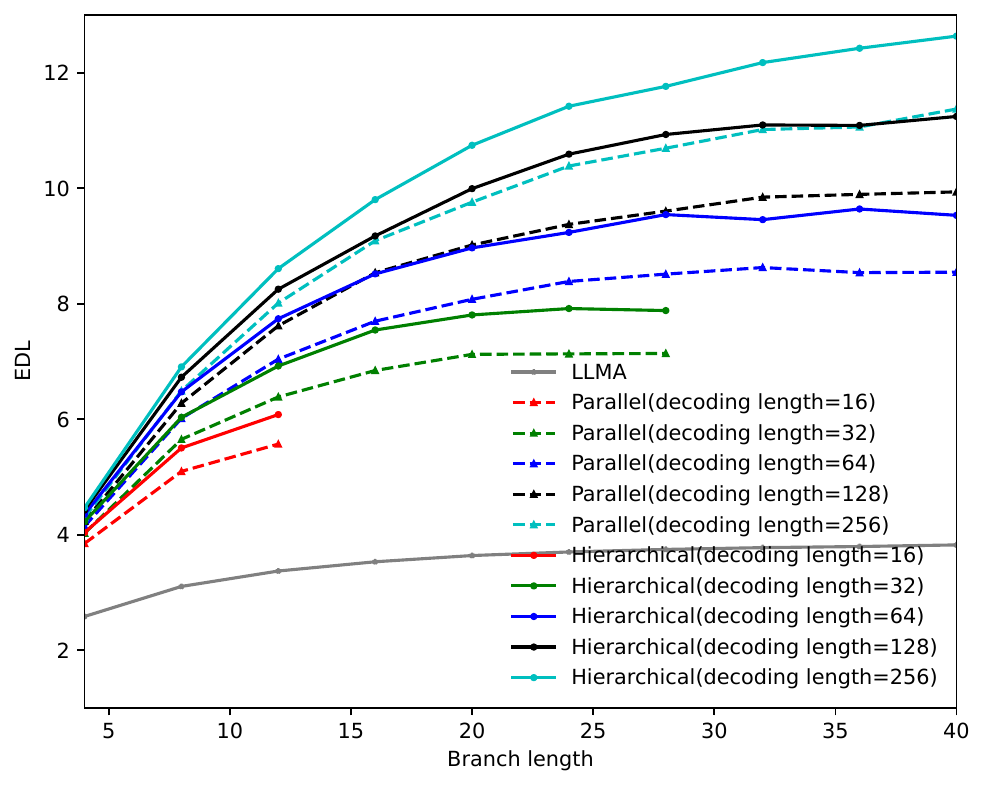}
	\caption{The decoding and branch length's impact on the \textbf{\textit{effective decoding length}}, \textbf{\textit{EDL}} using various accelerations.}
	\label{fig.edl}
\end{figure}

Generally speaking, it can be observed that as the decoding and branch lengths increase, there is an upward trend in the overall inference speed. As mentioned, with the single-branch strategy, LLMA is unable to further enhance the inference speed while its branch length surpasses 25. This limitation arises due to its inability to ensure a longer \textbf{\textit{EDL}}. By implementing the multi-branch strategy, \textit{Lookahead} is able to retrieve multiple branch drafts and thus guarantee a longer \textbf{\textit{EDL}} after the VA process, as illustrated in Figure \ref{fig.edl}. Consequently, this leads to a significantly higher inference speed, while maintaining the same branch length as LLMA in Figure \ref{fig.speed}. Furthermore, the improved \textbf{\textit{EDL}} as shown in Figure \ref{fig.edl} also provides more headroom for increasing the branch length and thereby enhancing the upper limit of the inference speed. 

In addition, Figure \ref{fig.edl} shows that being benefited from dealing with more branch drafts, \textit{Lookahead} using the hierarchical multi-branch draft promises the advantage of a longer \textbf{\textit{EDL}} and consequently improved inference speed, compared to using the parallel multi-branch draft with the identical branch and decoding lengths. It should be noted that in Figure \ref{fig.edl}, improving the decoding length promises a longer \textbf{\textit{EDL}} using \textit{Lookahead}, however the oversized decoding length that surpasses the \textbf{\textit{CDL}} fails to promise a higher inference speed, due to a more complex calculation logic for the tensor cores that has been introduced in the preliminary.

\subsubsection{Hyper-parameters in trie tree updating}

Table \ref{tbl.trie_update} showcases the inference speeds of \textit{Lookahead}, which exhibit variability depending on different procedures used for updating the trie tree. These procedures include branch inserting with prompt and/or output, branch eliminating, and node pruning. In comparison to the results obtained without a specific procedure, \textit{Lookahead} attains its optimal performance when all the aforementioned procedures are applied to dynamically update the trie tree, highlighting the necessity of these procedures in updating the trie tree.

\begin{table}[htb]
\caption{The inference speed of different updating procedures was evaluated, considering various conditions. In the evaluation, "W/o prompt" and "W/o output" refer to the absence of branching from the prompt and output, respectively. "W/o Pruning" signifies the exclusion of node count reduction in an oversized trie tree through pruning, "W/o eliminating" denotes the retention of branches derived from the prompt before processing the subsequent query.}

\small
\setlength\tabcolsep{1pt}
\label{tbl.trie_update}
\small
\begin{tabular}{cccccc}
	\toprule  
Condition & W/o prompt & W/o output & W/o pruning & W/o elimin & Lookahead \\ 
	\midrule  
Token/s  & 234.5 & 202.0 & 221.4 & 234.2 & 280.9 \\
\bottomrule 
\end{tabular}
\end{table}

% Intuitively, node capacity of trie tree will influence the inference performance. Considering this, we further empirically determine the trie tree's optimized capacity, as shown in Table \ref{tbl.cache_capacity}. Rather than using a fixed value, we normalized the node capacity relative to the decoding length to obtain a general hyper-parameter that would work well for various decoding lengths. Based on our findings, we determined that node capacities of 16 and 32 times the decoding length achieved the best performance. As a result, we have set the default capacity in our framework to be 16 times the decoding length.

The capacity of the trie tree has a significant impact on the acceleration performance. To determine the optimal capacity, we conduct empirical experiments, as outlined in Table \ref{tbl.cache_capacity}. Instead of using a fixed value, we normalize the node capacity in relation to the decoding length, which allows us to identify a suitable hyper-parameter that performs well across different decoding lengths. Based on our findings, we conclude that node capacities of 16 and 32 times the decoding length yield the best results. Consequently, we have set the default capacity in our framework to be 16 times the decoding length. We also examine the retrieving and updating time of trie tree with different capacity, the time is negligible when compared with the forward time.

\begin{table}[htb]
\caption{Inference Speed and retrieving/updating time of different capacity of trie tree. n*DL denotes the capicity is n times the decoding length.}
\small
\setlength\tabcolsep{2pt}
\label{tbl.cache_capacity}
\begin{tabular}{ccccccccc}
	\toprule  
Capacity  & 1*DL & 2*DL & 4*DL & 8*DL & 16*DL & 32*DL & 64*DL & 128*DL \\ 
	\midrule  
Token/s  & 254.5 & 258.7 & 268.8 & 273.3 & 280.7 & 280.8 & 279.5 & 278.3 \\
Retrieve(ms)  & 0.81 & 1.35 & 1.22 & 1.46 & 1.02 & 1.29 & 1.75 & 2.38 \\
Update(ms)  & 0.10 & 0.12 & 0.11 & 0.11 & 0.12 & 0.12 & 0.15 & 0.16 \\
\bottomrule 
\end{tabular}
\end{table}

\subsubsection{Lookahead's efficient memory usage}

% \begin{figure}[ht]
% 	\centering
% 	\includegraphics[width=3.0in]{figures/warmup.pdf}
% 	\caption{Inference Speed of different warming up sample counts. Our method achieves great improvement even without extra warmup samples, and performance gradually increases with larger warmup size.}
% 	\label{fig.warmup}
% \end{figure}

Figure \ref{tbl.cuda_memory} shows the peak GPU memory consumption for various decoding lengths (1, 2, 4, 8, 16, 32, 64, 128). The decoding length of 1 represents the experience where no \textit{Lookahead} is used, resulting in a GPU memory consumption of 20.25 GiB. Surprisingly, even with a decoding length of 128, the use of \textit{Lookahead} only leads to a negligible increase in GPU memory consumption of 0.6\%. This minimal increase in memory usage is practically negligible and has no significant impact on real-world applications.
Furthermore, we have also evaluated the CPU memory consumption of the trie tree. With the AntRAG dataset, the trie tree only utilizes a mere 260 MiB of memory. This value is negligible when compared to the total memory capacity of a mainstream server.

% \begin{figure}[htb]
% 	\centering
% 	\includegraphics[width=3.0in]{figures/cuda_memory.pdf}
% 	\caption{GPU Memory (i.e., device memory) of different decoding length. The GPU memory denotes the peak memory consumption in a inference process.}
% 	\label{fig.memory}
% \end{figure}

\begin{table}[htb]
\caption{Inference peak memory of different decoding lengths.}
\small
\setlength\tabcolsep{2pt}
\label{tbl.cuda_memory}
\begin{tabular}{ccccccccc}
\toprule  
Decoding length & 1 & 2 & 4 & 8 & 16 & 32 & 64 & 128 \\
\midrule  
Memory(GiB) & 20.25 & 20.27 & 20.27 & 20.27 & 20.27 & 20.29 & 20.38 & 20.39 \\
\bottomrule 
\end{tabular}
\end{table}

\subsection{Online Deployment}

Our framework has been widely used in real-world scenarios of Alipay since \textbf{April 2023}, due to its accessibility, accuracy-lossless and remarkable acceleration. To be concise, we only report 5 typical scenarios in Table \ref{tbl.online}. CHART2JSON is a scenario that uses a multi-modal model to convert a chart image to structured content with JSON format, it can achieve extraordinary acceleration due to plenty of template tokens in JSONs. Citizen biz agent, enterprise info QA and health suggestion are RAG scenarios, they are used for answering user questions with reference documents. Medical report summary is a scenario to summarize the content (texts after OCR) of a medical report image. With the assistance of our lookahead framework, the latency of all scenarios is decreased significantly while the generation results are the same as the original step-by-step strategy.

\begin{table}[htb]
\caption{Inference Speed of product scenarios. The speed is measured by mean latency (seconds) of a query.}
\small
\label{tbl.online}
\begin{tabular}{cccc}
\toprule  
Scenario & Baseline & Lookahead & Speedup \\ 
\midrule  
CHART2JSON & 12.97 & 2.07 & 6.26x \\
Citizen Biz Agent & 1.67 & 0.32 & 5.21x \\
Enterprise Info QA & 14.01 & 2.74 & 5.11x \\
Health Suggestion & 12.41 & 2.66 & 4.66x \\
Medical Report Summary & 3.33 & 1.25 & 2.66x \\
\bottomrule 
\end{tabular}
\end{table}

Meanwhile, we have integrated lookahead into other frameworks and obtained additional improvement. vLLM \cite{Kwon2023EfficientMM} is a widely used framework with state-of-the-art serving throughput, however we may suffer from its huge latency even when only one query is processing. We have implemented lookahead based on vLLM for the single-query situation and obtain 1.6 times acceleration on a real-life scenario about script generation for virtual human, shown in table \ref{tbl.vllm}.

\begin{table}[htb]
\caption{Inference Latency with Lookahead for vLLM.}
\small
\setlength\tabcolsep{2pt}
\label{tbl.vllm}
\begin{tabular}{ccccc}
\toprule  
Scenario  & vLLM  & vLLM+lookahead & speedup \\ 
\midrule  
Virtual Human Script & 5.109 & 3.203 & 1.60 \\
\bottomrule 
\end{tabular}
\end{table}

\section{Conclusion}
In our work, we empirically quantify that the main bottleneck of LLM inference is the IO bandwidth, rather than the FLOPs. Inspired by this, to take full advantage of the GPU's FLOPs redundancy, we innovatively develop \textit{Lookahead}, a generic framework that applies a hierarchical multi-branch draft strategy implemented with a trie tree to output more tokens per step than the traditional methods. We conduct extensive experiments and demonstrate that \textit{Lookahead} gains a substantial inference acceleration and cost reduction, with lossless generation accuracy. By simply adapting our framework to the latest LLMs, we have achieved a wide range of applications and promising prospects and will soon release our work with open source.

\clearpage

\bibliography{lookahead}

\clearpage

\appendix

\section{Workflow of lookahead}

We summarize the work flow of \textit{Lookahead} with pseudo code in the Algorithm \ref{algo}.

\begin{algorithm}
	%\textsl{}\setstretch{1.8}
	\renewcommand{\algorithmicrequire}{\textbf{Input:}}
	\renewcommand{\algorithmicensure}{\textbf{Output:}}
	\caption{Lookahead with multi-branch draft}
	\label{algo}
	\begin{algorithmic}[1]
            \REQUIRE decoding length $L_d$, branch length $L_b$, minimal count without re-retrieving $N_m$, node capacity of trie tree $N_{max}$, start token id $S$
            \ENSURE $O$
            \STATE initialization: trie tree $\mathcal{T}$
            \REPEAT
		\STATE initialization: output id list $O$=[$S$], KV cache $C_{kv}=[]$
		\STATE tokenize a query to token list $T$
            \FOR[branch inserting]{$i\leftarrow 1$ to len($T$)-1}  
            \STATE insert $T[i:i+L_b]$ into $\mathcal{T}$
            \IF{Node count of  $T$ > $N_m$ }
            \STATE do node pruning
            \ENDIF
            \ENDFOR
		\REPEAT 
            \FOR[trie tree retrieving]{$j \leftarrow len(O)$ to 1}
            \STATE obtain a prefix token list $T_p\leftarrow O[-j:]$
		\STATE match a sub trie tree $\mathcal{T}_s$ from $\mathcal{T}$ with $T_p$
            \IF{Node count of  $T_s$ < $N_m$ }
            \STATE continue
            \ENDIF
		\STATE select $L_d$-largest frequency nodes $N_{max}$ from $\mathcal{T}_s$
            \ENDFOR
            \STATE prepare token ids $T_{fp}$, position ids $P_{fp}$ and causal masks $M_{fp}$ for $N_{max}$
		\STATE get next tokens prediction via $LLM(C_{kv}, T_{fp},P_{fp},M_{fp)}$
            \STATE verify each branches and accept verified tokens $T_v$
            \STATE append $T_v$ to $O$
            \STATE incrementally put branches of $O$ to $\mathcal{T}$ 
            \STATE rearrange KV cache $C_{kv}$ with accepted tokens
		\UNTIL meet stopping criteria
            \STATE eliminate frequency of branches from the current prompt 
            \UNTIL all queries are procossed
	\end{algorithmic}
\end{algorithm}

\section{Dataset, Model and Device summary}

The experimental setup, including the datasets, models, and devices employed, is delineated in Table \ref{tbl.dataset}, Table \ref{tbl.model}, and Table \ref{tbl.device}, respectively. Specifically, the Llama2-7b-chat and Llama2-13b-chat models are utilized for the Dolly and GSM8k datasets, while the CodeLlama-7b and CodeLlama-13b models, which are fine-tuned on code dataset with Llama models, are applied to the HumanEval-x dataset. The prepossessed datasets are available in our repository.

\begin{table}[htb]
\caption{Summary of datasets.}
% \small
\setlength\tabcolsep{2pt}
\label{tbl.dataset}
\begin{tabular}{ccccc}
	\toprule  
	Dataset & Split & \#samples & \#tokens(prompt) & \#tokens(answer) \\
	\midrule  
	\multirow{2}{*}{AntRAG} & dev & 7,605 & 243.1 & 81.9 \\
             & test & 1,000 & 241.0 & 82.0 \\
             \midrule  
	\multirow{2}{*}{Dolly} & dev & 13,850 & 298.3 & 101.8 \\
             & test & 1,000 & 301.5 & 104.8 \\
             \midrule  
	\multirow{2}{*}{GSM8k} & dev & 7,792 & 66.9 & 130.8 \\
             & test & 1,000 & 67.9 & 131.9 \\
             \midrule  
	\multirow{2}{*}{HumanEval-x} & dev & 410 & 187.0 & 125.9 \\
             & test & 410 & 139.8 & 82.1 \\
	\bottomrule 
\end{tabular}
\end{table}

\begin{table}[htb]
\caption{Summary of models.}
% \small
% \setlength\tabcolsep{2pt}
\label{tbl.model}
\begin{tabular}{cccc}
\toprule  
Model & AntGLM-10B  & Llama-7B & Llama-13B \\
\midrule  
Params & 10.14B & 6.74B & 12.71B \\
Vocab size & 115328 & 32000 & 32000 \\
Layer & 48 & 32& 40 \\
Hidden size & 4096 & 4096 & 5120 \\
Attention head & 32 & 32 & 40 \\
MLP size & 16384 & 11008 & 13824 \\
\bottomrule 
\end{tabular}
\end{table}

\begin{table}[htb]
\caption{Summary of devices.}
% \small
% \setlength\tabcolsep{2pt}
\label{tbl.device}
\begin{tabular}{cccc}
\toprule  
Device & A100-80G  & A10 & V100-32G \\
\midrule  
Memory & 80G HBM2e & 24G GDDR6 & 32G HBM2 \\
Memory bandwidth & 2,039GB/s & 600GB/s & 900GB/s \\
FLOPs(FP16) & 312T & 125T & 125T \\
\bottomrule 
\end{tabular}
\end{table}

\section{Prompt templates.}

Similar to \cite{alpaca}, we construct a prompt for the Dolly dataset with the template in Table \ref{tbl.prompt}.

\begin{table}[ht]
\caption{Prompt templates of Dolly.}
\small
\label{tbl.prompt}
\begin{tabular}{cc}
\toprule  
Prompt Type & Propmt Template \\ 
\midrule  
W/ reference & \begin{tabular}{l}
    Below is an instruction that describes a task, paired \\ with an input that provides further context. Write \\ a response that appropriately completes the request. \\ \#\#\# Instruction:  \{instruction\}   \\ \#\#\# Input: \{reference\}   \\ \#\#\# Response:  \\
\end{tabular}  \\ 
\midrule  
W/o reference & \begin{tabular}{l}
    Below is an instruction that describes a task. Write \\ a response that appropriately completes the request.   \\ \#\#\# Instruction: \{instruction\} \\  \#\#\# Response:  \\
\end{tabular} \\

\bottomrule 
\end{tabular}
\end{table}

\section{Warmup Analysis}

\begin{figure}[htb]
	\centering
	\includegraphics[width=3.0in]{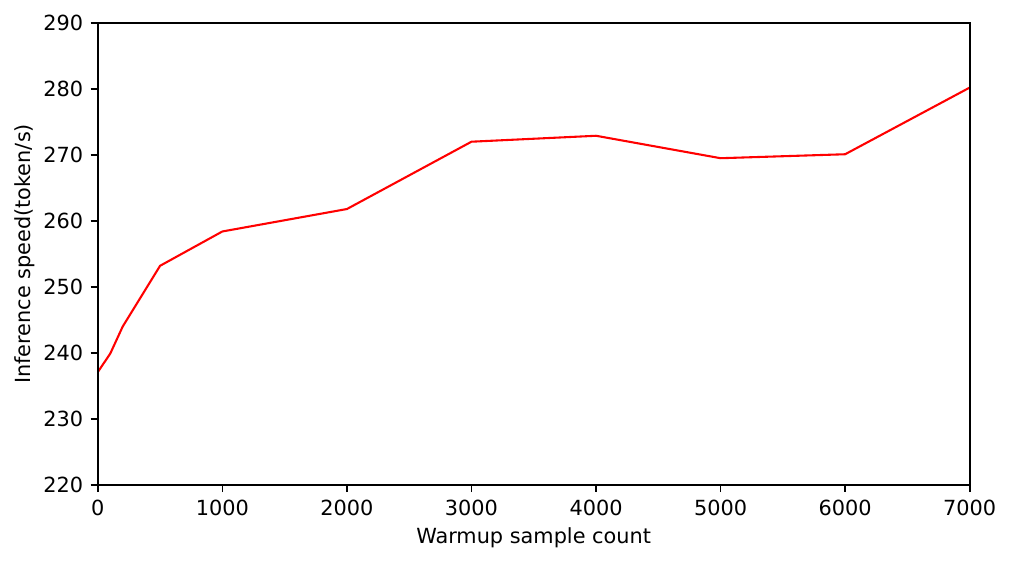}
	\caption{Inference speed with various warm-up sample count.}
	\label{fig.warmup}
\end{figure}

To accurately measure inference speed, a warm-up strategy is employed where responses from the development set are preloaded into the trie tree. As illustrated in Figure \ref{fig.warmup}, inference speed is positively correlated with the number of warm-up samples. Given that a model may process hundreds of thousands of requests over its lifespan, it is reasonable to infer that the average speed would exceed that observed during the initial thousands of requests.

\section{Batch Inference}

\begin{figure}[htb]
	\centering
	\includegraphics[width=3.0in]{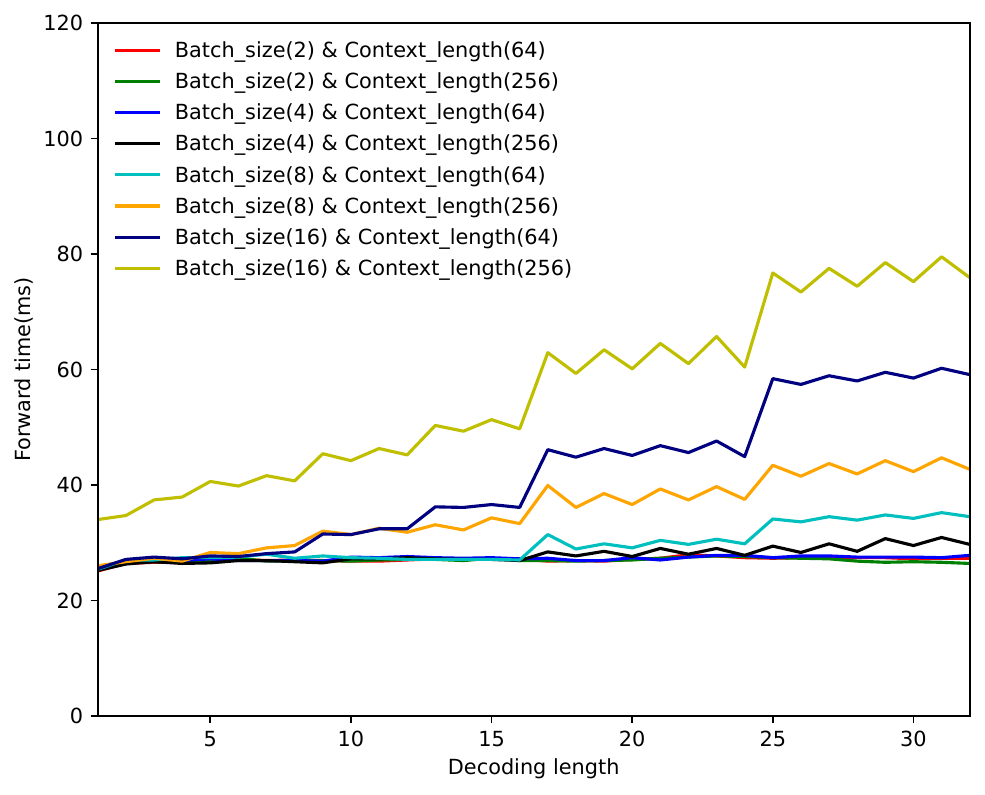}
	\caption{GPU Memory of various batch size, context length and decoding length.}
	\label{fig.antglm_batch_perf}
\end{figure}

The experiments described were conducted under single-query conditions (i.e., a batch size of 1). It is essential to recognize that batch inference (i.e., batch size > 1) incurs greater computational demand than single-query inference. To ascertain the computational threshold for batch inference, we evaluated the forward time across varying batch sizes, context lengths, and decoding lengths, as depicted in Figure \ref{fig.antglm_batch_perf}. It can be seen that while the batch size is 4 for instance, an increase in decoding length from 5 to 32 does not substantially augment the forward time. Moreover, forward time still remains manageable for a decoding length increasing from 5 to 16 while the batch size is 16. The GPUs still show their redundancies in these scenarios and lookahead in batch inference could still be applied as an effective approach.
% The results indicate that an increase in decoding length to 32 does not substantially augment the forward time for batch sizes up to 4, and forward time remains manageable for a decoding length of 16 even at a batch size of 16. This suggests that implementing lookahead in batch inference could be a viable approach.

To implement \textit{lookahead} within batch inference parameters, one must navigate variances in KV cache sequence length and attention masks, with implementation specifics available in the repository. To our knowledge, this constitutes the first batch implementation of speculative decoding methodologies. As evidenced in Table \ref{tbl.batch_infer}, \textit{lookahead} retains its efficacy even with non-unitary batch sizes, demonstrating a notable increase in inference speed relative to traditional step-by-step and LLMA-based batch inference approaches. Despite the less pronounced speedup compared to unitary batch scenarios, two primary factors contribute to this: first, large batch sizes decrease the GPU's redundancy, particularly with extensive context and decoding lengths, thereby capping potential speed enhancements. Second, the KV cache lengths' heterogeneity within a batch introduces extra computational overhead in attention operations, diminishing inference velocity.

Nonetheless, it is critical to emphasize that Lookahead is tailored for time-sensitive environments rather than those sensitive to throughput. Future work includes optimizing throughput by incorporating continuous batching \cite{Yu2022OrcaAD} and high-efficiency attention mechanisms \cite{dao2022flashattention, Kwon2023EfficientMM}.

\begin{table}[htb]
\caption{Inference Speed of different batch sizes with the AntRAG dataset.}
\small
\label{tbl.batch_infer}
\begin{tabular}{ccccc}
\toprule  
\multirow{2}{*}{Method} & \multicolumn{2}{c}{Batch size=2} & \multicolumn{2}{c}{Batch size=4}\\ 
\cline{2-5}
 & token/s & speedup & token/s & speedup \\ 
\midrule  
Baseline & 68.0 & 1.00x & 88.2 & 1.00x \\
LLMA & 185.2 & 2.72x & 214.0 & 2.43x \\
% Lookahead(Parallel) & 274.4 & 4.04x & 291.2 & 3.30x \\
Lookahead & 285.5 & 4.20x & 299.5 & 3.40x \\
\bottomrule 
\end{tabular}
\end{table}

% \subsection{Continuous Batch}

% \textcolor[rgb]{1,0,0}{We also enhance other framework with Lookahead. We also use lookahead to further improve performance for inference with continuous batch mechanism. Note that we currently only partly implement lookahead with the two frameworks, with just one branch draft, fully implementation is on working.}

% \begin{table}[htb]
% \caption{Inference Latency with Lookahead.}
% \small
% \setlength\tabcolsep{2pt}
% \label{tbl.cache_capacity}
% \begin{tabular}{ccccccccc}
% 	\toprule  
% Method  & Latency(s) & throughput(token/s) \\ 
% 	\midrule  
% continuous batch & 0.565 & 33.7  \\
% continuous batch+lookahead & 0.508(x1.11) & 42.4(x1.26)  \\
% \bottomrule 
% \end{tabular}
% \end{table}

\end{document}